\documentclass[acmsmall]{acmart}

\AtBeginDocument{%
  \providecommand\BibTeX{{%
    \normalfont B\kern-0.5em{\scshape i\kern-0.25em b}\kern-0.8em\TeX}}}

\setcopyright{rightsretained}
\acmConference[Contestability'19]{2019 CSCW Workshop on Contestability in Algorithmic Systems}{November
9--13}{Austin, TX, USA}
\acmDOI{}
\acmISBN{}

\begin{document}

\title{Designing for Contestation:\\Insights from Administrative Law}

\author{Henrietta Lyons}
\email{hlyons@student.unimelb.edu.au}
\orcid{1234-5678-9012}
\affiliation{%
  \institution{The University of Melbourne}
  \streetaddress{Parkville Campus}
  \city{Melbourne}
  \state{Victoria}
  \postcode{3010}
}

\author{Eduardo Velloso}
\email{eduardo.velloso@unimelb.edu.au}
\orcid{TBC}
\affiliation{%
  \institution{The University of Melbourne}
  \streetaddress{Parkville Campus}
  \city{Melbourne}
  \state{Victoria}
  \postcode{3010}
}

\author{Tim Miller}
\email{tmiller@unimelb.edu.au}
\orcid{TBC}
\affiliation{%
  \institution{The University of Melbourne}
  \streetaddress{Parkville Campus}
  \city{Melbourne}
  \state{Victoria}
  \postcode{3010}
}


\maketitle

\section{Introduction}
Algorithmic decision-making systems are increasingly being deployed to make, or to support humans to make, decisions that impact people's lives in significant ways. Yet, {\itshape decision subjects}, those affected by algorithmic decisions, can be limited in their ability to contest these decisions. For example, the Education Value-Added Assessment System (EVAAS), a statistical method used to predict academic growth, was used by the Houston Independent School District to evaluate teachers' performance and, in a number of cases, to terminate teachers' contracts. Twelve teachers and the Houston Federation of Teachers successfully argued in court that the teachers' constitutional right to due process was violated because they were unable to contest, or `meaningfully challenge', the termination of their contracts because there was a `lack of sufficient information' --- the private company that designed EVAAS would not release the source codes or methodology used as they were proprietary trade secrets \cite{Houstonteachers17}.

Even for decision subjects who are able to understand why a decision has been made and are provided with means to contest that decision, contestation systems can be seen as severely lacking \cite{MyersWest2018}. Sarah Myers West studied content moderation across a number of social media platforms, most of which offered a way for users to contest a decision to remove their content from the platform \cite{MyersWest2018}. Myers West reported user dissatisfaction with the contestation systems for a number of reasons, including a lack of clear instruction about how to lodge an appeal, no reply being received, no resolution being reached after a challenge has been lodged, and a lack of access to human intervention.

These examples demonstrate that being able to challenge algorithmic decisions is important to decision subjects, yet numerous factors can limit a person's ability to contest such decisions. We propose that administrative law systems, which were created to ensure that governments are kept accountable for their actions and decision making in relation to individuals \cite{Bannister18, Policy}, can provide guidance on how to design contestation systems for algorithmic decision-making. 

There are similarities between government decision-making and algorithmic decision-making that suggest there is value in considering how the administrative law system enables contestation. For example, in both cases decision making can be said to occur `behind closed doors', which limits transparency, raises questions about accountability, and has the potential to impact public trust. In this paper, we focus on the specific case of the Australian administrative law system because this is the authors' country of residence, and because of the lead author's familiarity with the Australian legal system and her experience working within a variety of government departments and agencies within Australia. Further to this, the Australian administrative law system has been refined over the past 40 years: it provides a comprehensive and well-designed example of an established contestation system.

This paper contributes: (1) a methodological proposal for the development of a framework that can be used to guide the design of contestation systems for algorithmic decision-making, based on the Australian administrative law system; (2) a summary of key features of the Australian administrative law system that enable individuals to contest government decision-making; (3) a list of considerations to help to prompt design thinking in relation to contestation systems for algorithmic decision-making. While this paper raises more questions than it answers, these questions form a useful base from which to consider the development of contestation systems that allow decision subjects to meaningfully challenge algorithmic decisions.

\section{Related Work}

\subsection{Designing for contestability}
Recent work highlights the benefits of designing `contestability' into machine learning systems used by experts for decision support \cite{Hirsch17}. We view `designing for contestation' as quite distinct from `designing for contestability'. 
 `Designing for contestability' enables expert users to engage interactively with a machine learning system, to explore it, and to provide it with feedback, with the aim to produce better decisions \cite{Hirsch17}. In contrast, `designing for contestation' does not (necessarily) entail an interactive design component within the algorithmic decision-making system. Indeed, it may not be appropriate for a decision subject to interact and engage with the algorithmic system themselves, depending on the decision. The aim of designing for contestation is to enable decision subjects to appeal a decision by providing them with the means to do so. For example, as highlighted by the EVAAS court case \cite{Houstonteachers17}, by providing enough information about the decision to allow decision subjects to `meaningfully challenge' it.  Or, by ensuring that an effective process to appeal a decision exists.  

\subsection{Contesting algorithmic decisions}
In a 2017 workshop paper on algorithmic appeals, Vaccaro and Karahalios highlight that established contestation systems can be used to inform contestation design for algorithmic decision-making systems \cite{Vaccaro17}. Vaccaro and Karahalios explored three contexts where people can currently seek to review decisions (the court system, credit scoring, and insurance claims) to understand how these systems enable contestation and to reveal challenges that algorithmic appeal systems are likely to face. The authors argue that being able to contest a decision is not only about understanding, or being able to interpret, the decision made by an algorithm, although this is certainly a necessary element---there is also a need for a system to support appeals. 

Myers West surveyed 519 social media platform users to understand their experience with content modification systems on the platforms that they use \cite{MyersWest2018}. Myers West found that a majority of the platforms allowed users to appeal decisions to remove content, but that users who chose to make an appeal (n=230) experienced difficulties with the process. For example, users reported that while the platform notified them of the option to appeal, they were not provided with any instructions about how to go about it. Other users were frustrated by the lack of response or resolution offered by the platform after they had lodged a complaint, and the lack of human intervention that was provided. Of those surveyed, Myers West reports that 245 users chose not to appeal, with some explaining that they did not know how to lodge an appeal or that they chose not to because they did not think that they would receive a response. 

Almada considers algorithmic decision contestation in the context of the General Data Protection Regulation, in which Article 22(3) provides a right to contest a decision and a right of human intervention \cite{Almada19}. Almada argues that `[a]t a bare minimum, a system that is contestable by design should be built in ways that allow users and third parties to effectively seek human intervention in a given automated decision', and he suggests that literature relating to privacy by design and designing for contestability can provide insights into designing for contestation.

\subsection{Explainable artificial intelligence}
While algorithmic decision-making has been occurring for decades, in recent years advances in machine learning has resulted in an increased use of these often ``black box'' systems to support, or to make, high consequence decisions \cite{Abdul18, Ribera19, Wang19}. As a result, there have been calls for these algorithms to be made explainable, transparent, accountable, and fair to ensure that users are able to trust their decisions \cite{Abdul18, Ribeiro16, Ribera19}. While these calls have resulted in the burgeoning research field of explainable artificial intelligence (XAI), limited attention has been given to contestation as a way to address these goals. There is potential here for XAI to be informed by contestation. But in addition, there is a real opportunity for contestation to be supported by XAI.

XAI is a dynamic field: technical solutions to produce explanations are being proposed, and a greater focus is now also being placed on the need for human-centred explanation design \cite{miller17, Abdul18, Wang19, Adadi18}, with more research being conducted into what users require from an explanation \cite{Ribera19}. An explanation would provide decision subjects with valuable information that could inform a basis from which they could contest a decision. Yet, few papers have proposed helping a decision subject to contest a decision as a goal for an explanation. Notable exceptions to this include Wachter et al., who suggest that a potential goal for an explanation is `to provide information that helps contest automated decisions when an adverse or otherwise undesired decision is received', \cite{Watcher18} and Binns et al. who studied the affect of explanation style on justice perceptions, but did not delve into contestation per se \cite{Binns18}.

\section{The Australian administrative law system}
\begin{figure}[h]
  \centering
  \includegraphics[width=\linewidth]{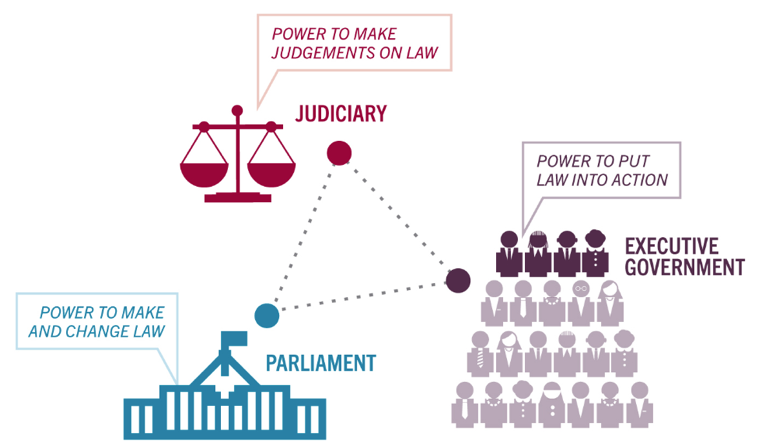}
  \caption{The parliament, executive and judiciary each have a separate role to play in Australian governance [Public domain], via The Parliamentary Education Office (\url{https://www.peo.gov.au/understand-our-parliament/how-parliament-works/separation-of-powers/separation-of-powers/}).}
  \Description{The figure indicates how the separation of powers works in Australia: it contains reference to the judiciary which has the power to make judgements on law, reference to the executive government, which has the power to put law into action, and the parliament, which has the power to make and change the law}
  \label{separation_of_powers}
\end{figure}

We propose that administrative law systems can provide guidance on how to design contestation systems for algorithmic decision-making. In this paper, we focus on the Australian administrative law system, which regulates commonwealth government decision-making that affects individuals \cite{Policy}. This system enables decision subjects to contest decisions made by decision-makers in the executive arm of the government --- namely officers in government departments. In Australia, the role of the executive government is to administer the laws created by parliament \cite{ParlEdu}. Figure~\ref{separation_of_powers} sets out the roles of the parliament, the executive, and the judiciary. In its role administering and applying laws, the executive has a direct impact on individual members of the public: it has the power to make decisions that affect a person's rights, including whether a person is entitled to welfare, is granted a visa, or is granted a licence \cite{Handbook}. Bannister, Olijnyk, and McDonald state that `[t]he executive's application of laws affects the day-to-day lives of individuals more often and more directly than the actions of the legislative and judicial branches of government' \cite{Bannister18}.

Australian administrative law is made up of a number of elements that each play a role in regulating the actions of the executive: it is a system of accountability. Bannister, Olijnyk, and McDonald describe in detail the elements of the system, which include:
\begin{itemize}
\item Mechanisms that enhance transparency by allowing access to information, for example through rules enabling freedom of information;
\item A range of bodies that can carry out investigations relating to the actions or decisions of the executive, including the Commonwealth Ombudsman, the Administrative Review Council, and Royal Commissions; and
\item Mechanisms that provide individuals with avenues to contest decisions that affect them, for example via an internal or external review \cite{Bannister18}. 
\end{itemize}
While the last element listed above is the most informative for developing a contestation framework, understanding that administrative law is a system is important. There are numerous ways for the executive to breach its powers, and the administrative law mechanisms provide different avenues to hold the executive to account. For example, the contestation system supports individual rights and provides individual redress, whereas bodies such as the Ombudsman can investigate systemic issues with decision making that may not be identifiable at the individual level. In a similar vein, in the context of algorithmic decision-making, providing a way for decision subjects to contest decisions will enable the review of individual decisions, but may not reveal systemic problems embedded in an algorithm, such as biases that result in discrimination. An auditing program may better suit that task by enabling third-parties to measure how well the algorithm performs against predefined goals, which could include fairness and equity \cite{BENNETT18, Guszcza18, Vaccaro17}. When designing a contestation system, designers need to understand what that system can and cannot feasibly achieve.

\section{Designing for contestation}

We are in the process of developing a framework that can be used to guide the design of contestation systems for algorithmic decision-making. The first column in Table 1 contains a list of the key features of the Australian administrative law system that enable individuals to contest decisions made by the government that affect them. An informal approach was taken to compile this initial list of key features: the lead author drew on her legal training and experience working within government agencies to select the pertinent features. Criteria used to select these features was whether the feature limits, or supports, a decision subject to contest a decision. The features in column 1 will be augmented and refined with further research into the administrative law system as well as other established contestation systems. 

The second column of Table 1 contains questions and considerations derived from the first column that will be useful to consider when designing contestation systems for algorithmic decision-making. Again, an informal approach was taken to develop these questions; they were the product of brainstorming prompted by reflection on the features in column 1. The questions contained in the second column are not exhaustive, they can (and should) be added to and refined.

Many jurisdictions, including the United States, the United Kingdom, Canada, and France have administrative law systems that govern the actions and decision-making of their respective governments, however the specific mechanisms and processes used to meet this aim differ. For example, while `fair hearing' forms part of the Australian administrative law system, in the United States, the equivalent right is constitutional: a citizen is entitled to due process of law if being deprived of life, liberty or property. 

\begin{table}
  \caption{A framework to consider when designing contestation systems for algorithmic decision-making}
  \label{tab:freq}
  \begin{tabular}{p{6.5cm}p{6.5cm}}
 \toprule
    Pertinent features of the administrative law system that enable individual contestation & Considerations for contestation systems for algorithmic decision-making\\
    \midrule
    \texttt Not all decisions are reviewable: a person needs `standing' to appeal a decision e.g. they need to have been `affected' by the decision; the Administrative Appeals Tribunal can only review decisions if legislation provides that power. & Should there be limits on who can contest a decision? Which decisions should be able to be contested? All decisions? Only decisions that affect certain rights or interests? \\
 \space\\
    \texttt Best practice is to notify a person if the decision is likely to be adverse and to give them an opportunity to respond. This helps to ensure that a person is given a `fair hearing'. & Is a `fair hearing' required? What process would provide decision subjects with a fair hearing'?\\
\space\\
    \texttt Once a decision is made, notice of that decision is provided to the person along with information about how to initiate a review. & How will a person be notified of a decision? What information will be provided about review? How should an interface for contestation be designed?\\
    \space\\
    \texttt When a decision subject has standing to appeal a decision, they can request a `statement of reasons', which should contain: findings on material questions of fact; evidence or other material the findings were based on; and the reasons for the decision. & What does a statement of reasons look like for an algorithm? What information does a decision subject need to be able to meaningfully contest a decision? Can this information be provided? What if the algorithm is a "black box"? \\
    \space\\
    \texttt Various avenues of review are available e.g. internal merits review, external merits review, and judicial review. There are processes dictating how to seek each type of review. Internal merits review (where the decision making agency reviews the decision) is usually the first step.  & What types of review can be provided? If a decision subject can access an internal review, who will carry out the review? A human or an algorithm? Would an algorithmic review simply result in the same decision? Will further avenues of review be provided?\\
    \space\\
    \texttt The review bodies are limited in their ability to review a decision e.g. a merits review can determine whether the `right' decision was made by looking at the relevant facts, law, and policy, whereas a judicial review can only consider whether the decision-making process was lawful. & What aspects of a decision or decision-making process can be contested? Just the decision/output that pertains to the decision subject? The training data? The inputs? The algorithm's decision rules? The process for deriving the decision-making model?\\
    \space\\
    \texttt The remedies provided to a decision subject differ depending on the review body e.g. a merits review can make a fresh decision, whereas a judicial review cannot make a new decision, but can order that a new decision be made by the original decision maker. & What redress can be provided to a decision subject? Can a new decision be made? How will it be made? If it is determined that an input was used in error, can a new decision be made without considering that input?\\

  \bottomrule
\end{tabular}
\end{table}

\section{Discussion}

The draft framework in Table 1 displays key elements that make up an established contestation system. Given its legal nature, the administrative law system is heavily rule-based and quite complex. Algorithmic decision-making contestation systems may not need to be as complicated as this. However, when decisions are being made that affect a person's rights and interests, a comprehensive and considered approach to contestation may be required.
A one-size-fits-all approach to designing a contestation system will not be appropriate given the range of decisions being made and influenced by algorithmic decision-making systems, and the various contexts these decisions are being made in. In addition to understanding user needs, the use case, and constraints, designers must be alive to the legal and regulatory contexts in which decisions are being made, to ensure that legal requirements are met. For example, Article 22(3) of the General Data Protection Regulations creates a right to contest certain automated decisions; and there is a constitutional right to due process in the United States.

Further, given that the administrative law system is designed to regulate human decision-makers, some of the elements in column 1 of Table 1 may not be relevant for algorithmic decision-making. Similarly, there may be additional, unique elements that are needed to cater for algorithmic decision-making. For example, a major challenge presented by "black box" algorithmic decision-making systems is that their opacity obscures the decision-making process. Thus, it is unlikely to be easy for an algorithmic decision-making system, or a person using such a system, to produce the same information contained in a `statement of reasons' (which is an element of the Australian administrative law system). However, with advances in XAI research, methods are being developed to produce explanations for decisions. An explanation that provides information about the decision and the decision-making process would provide decision subjects with a base from which they can meaningfully contest a decision that has been made, or influenced, by an algorithmic decision-making system. In fact, depending on the type of information that a decision subject wishes to contest, understanding this internal logic of an algorithm may not even be required; for example, counterfactual explanations can be used to provide valuable information to decision subjects without opening the "black box" \cite{Watcher18}. Further research needs to be conducted into what type of information would allow a decision subject to meaningfully contest an algorithmic decision, and how this could be produced. 

\section{Future work}
We will continue to develop the framework to guide the design of contestation systems for algorithmic decision-making by taking into account other contestability systems and by  refining and augmenting the key features using formal methodological approaches such as thematic coding and content expert interviews. We also propose to conduct research into user requirements for contestation systems, and in particular into what type of information users require in order to contest an algorithmic decision.

\section{Conclusion}

For decision subjects who have been affected by algorithmic decision-making, such as the Houston teachers whose contracts were terminated, being able to challenge an adverse algorithmic decision is vital. Yet, contestation systems for algorithmic decisions are currently underdeveloped, with decision subjects left with limited information about how the decision was made and how to contest the decision. Existing contestation systems, such as administrative law systems, can provide valuable insights into how contestation systems for algorithmic decision-making can be designed so that decision subjects can meaningfully contest decisions that affect them.

\begin{acks}
Henrietta Lyons's work is supported by the Australian Government Research and Training Program scholarship and a Google Travel Scholarship. This research was partly funded by Australian Research Council Discovery Grant DP190103414 \emph{Explanation in Artificial Intelligence: A Human-Centred Approach}. Eduardo Velloso is the recipient of an Australian Research Council Discovery
Early Career Researcher Award (Project Number: DE180100315) funded by the Australian Government.
\end{acks}

\bibliographystyle{ACM-Reference-Format}
\bibliography{sample-base}


\begin{thebibliography}{19}


\ifx \showCODEN    \undefined \def \showCODEN     #1{\unskip}     \fi
\ifx \showDOI      \undefined \def \showDOI       #1{#1}\fi
\ifx \showISBNx    \undefined \def \showISBNx     #1{\unskip}     \fi
\ifx \showISBNxiii \undefined \def \showISBNxiii  #1{\unskip}     \fi
\ifx \showISSN     \undefined \def \showISSN      #1{\unskip}     \fi
\ifx \showLCCN     \undefined \def \showLCCN      #1{\unskip}     \fi
\ifx \shownote     \undefined \def \shownote      #1{#1}          \fi
\ifx \showarticletitle \undefined \def \showarticletitle #1{#1}   \fi
\ifx \showURL      \undefined \def \showURL       {\relax}        \fi
\providecommand\bibfield[2]{#2}
\providecommand\bibinfo[2]{#2}
\providecommand\natexlab[1]{#1}
\providecommand\showeprint[2][]{arXiv:#2}

\bibitem[\protect\citeauthoryear{??}{Hou}{2017}]%
        {Houstonteachers17}
 \bibinfo{year}{2017}\natexlab{}.
\newblock \bibinfo{title}{Houston Federation of Teachers, Local 2415, et al v
  Houston Independent School District 251 F.Supp.3d 1168}.
\newblock
\newblock


\bibitem[\protect\citeauthoryear{Abdul, Vermeulen, Wang, Lim, and
  Kankanhalli}{Abdul et~al\mbox{.}}{2018}]%
        {Abdul18}
\bibfield{author}{\bibinfo{person}{Ashraf Abdul}, \bibinfo{person}{Jo
  Vermeulen}, \bibinfo{person}{Danding Wang}, \bibinfo{person}{Brian~Y. Lim},
  {and} \bibinfo{person}{Mohan Kankanhalli}.} \bibinfo{year}{2018}\natexlab{}.
\newblock \showarticletitle{Trends and Trajectories for Explainable,
  Accountable and Intelligible Systems: An HCI Research Agenda}. In
  \bibinfo{booktitle}{\emph{Proceedings of the 2018 CHI Conference on Human
  Factors in Computing Systems}} \emph{(\bibinfo{series}{CHI '18})}.
  \bibinfo{publisher}{ACM}, \bibinfo{address}{New York, NY, USA}, Article
  \bibinfo{articleno}{582}, \bibinfo{numpages}{18}~pages.
\newblock
\showISBNx{978-1-4503-5620-6}
\urldef\tempurl%
\url{https://doi.org/10.1145/3173574.3174156}
\showDOI{\tempurl}


\bibitem[\protect\citeauthoryear{Adadi and Berrada}{Adadi and Berrada}{2018}]%
        {Adadi18}
\bibfield{author}{\bibinfo{person}{Amina Adadi} {and} \bibinfo{person}{Mohammed
  Berrada}.} \bibinfo{year}{2018}\natexlab{}.
\newblock \showarticletitle{Peeking Inside the Black-Box: A Survey on
  Explainable Artificial Intelligence (XAI)}.
\newblock \bibinfo{journal}{\emph{IEEE Access}}  \bibinfo{volume}{6}
  (\bibinfo{year}{2018}), \bibinfo{pages}{52138--52160}.
\newblock
\urldef\tempurl%
\url{https://doi.org/10.1109/ACCESS.2018.2870052}
\showDOI{\tempurl}


\bibitem[\protect\citeauthoryear{Almada}{Almada}{2019}]%
        {Almada19}
\bibfield{author}{\bibinfo{person}{Marco Almada}.}
  \bibinfo{year}{2019}\natexlab{}.
\newblock \showarticletitle{Human Intervention in Automated Decision-making:
  Toward the Construction of Contestable Systems}. In
  \bibinfo{booktitle}{\emph{Proceedings of the Seventeenth International
  Conference on Artificial Intelligence and Law}} \emph{(\bibinfo{series}{ICAIL
  '19})}. \bibinfo{publisher}{ACM}, \bibinfo{address}{New York, NY, USA},
  \bibinfo{pages}{2--11}.
\newblock
\showISBNx{978-1-4503-6754-7}
\urldef\tempurl%
\url{https://doi.org/10.1145/3322640.3326699}
\showDOI{\tempurl}


\bibitem[\protect\citeauthoryear{Attorney-General's~Department}{Attorney-General's~Department}{2011}]%
        {Policy}
\bibfield{author}{\bibinfo{person}{Australian~Government
  Attorney-General's~Department}.} \bibinfo{year}{2011}\natexlab{}.
\newblock \bibinfo{title}{Australian Administrative Law Policy Guide}.
\newblock
\newblock
\urldef\tempurl%
\url{https://www.ag.gov.au/LegalSystem/AdministrativeLaw/Documents/Australian-administrative-law-policy-guide.pdf}
\showURL{%
Retrieved October 2, 2019 from \tempurl}


\bibitem[\protect\citeauthoryear{Bannister, Olijnyk, and McDonald}{Bannister
  et~al\mbox{.}}{2018}]%
        {Bannister18}
\bibfield{author}{\bibinfo{person}{Judith Bannister}, \bibinfo{person}{Anna
  Olijnyk}, {and} \bibinfo{person}{Stephen McDonald}.}
  \bibinfo{year}{2018}\natexlab{}.
\newblock \bibinfo{booktitle}{\emph{Government accountability: Australian
  administrative law} (\bibinfo{edition}{2nd.} ed.)}.
\newblock \bibinfo{publisher}{Cambridge University Press},
  \bibinfo{address}{Port Melbourne, Vic}.
\newblock


\bibitem[\protect\citeauthoryear{Bennett}{Bennett}{2018}]%
        {BENNETT18}
\bibfield{author}{\bibinfo{person}{Tess Bennett}.}
  \bibinfo{year}{2018}\natexlab{}.
\newblock \bibinfo{title}{Governments Should Independently Audit AI Tools For
  Fairness: Analytics Expert}.
\newblock
\newblock
\urldef\tempurl%
\url{https://which-50.com/governments-should-independently-audit-ai-tools-for-fairness-analytics-expert/}
\showURL{%
Retrieved October 3, 2019 from \tempurl}


\bibitem[\protect\citeauthoryear{Binns, Van~Kleek, Veale, Lyngs, Zhao, and
  Shadbolt}{Binns et~al\mbox{.}}{2018}]%
        {Binns18}
\bibfield{author}{\bibinfo{person}{Reuben Binns}, \bibinfo{person}{Max
  Van~Kleek}, \bibinfo{person}{Michael Veale}, \bibinfo{person}{Ulrik Lyngs},
  \bibinfo{person}{Jun Zhao}, {and} \bibinfo{person}{Nigel Shadbolt}.}
  \bibinfo{year}{2018}\natexlab{}.
\newblock \showarticletitle{It's Reducing a Human Being to a Percentage}.
\newblock \bibinfo{journal}{\emph{Proceedings of the 2018 CHI Conference on
  Human Factors in Computing Systems - CHI '18}} (\bibinfo{year}{2018}).
\newblock
\showISBNx{9781450356206}
\urldef\tempurl%
\url{https://doi.org/10.1145/3173574.3173951}
\showDOI{\tempurl}


\bibitem[\protect\citeauthoryear{Guszcza, Rahwan, Bible, Cebrian, and
  Katyal}{Guszcza et~al\mbox{.}}{2018}]%
        {Guszcza18}
\bibfield{author}{\bibinfo{person}{James Guszcza}, \bibinfo{person}{Iyad
  Rahwan}, \bibinfo{person}{Will Bible}, \bibinfo{person}{Manuel Cebrian},
  {and} \bibinfo{person}{Vic Katyal}.} \bibinfo{year}{2018}\natexlab{}.
\newblock \bibinfo{title}{Why We Need to Audit Algorithms}.
\newblock
\newblock
\urldef\tempurl%
\url{https://hbr.org/2018/11/why-we-need-to-audit-algorithms}
\showURL{%
Retrieved October 3, 2019 from \tempurl}


\bibitem[\protect\citeauthoryear{Hirsch, Merced, Narayanan, Imel, and
  Atkins}{Hirsch et~al\mbox{.}}{2017}]%
        {Hirsch17}
\bibfield{author}{\bibinfo{person}{Tad Hirsch}, \bibinfo{person}{Kritzia
  Merced}, \bibinfo{person}{Shrikanth Narayanan}, \bibinfo{person}{Zac~E.
  Imel}, {and} \bibinfo{person}{David~C. Atkins}.}
  \bibinfo{year}{2017}\natexlab{}.
\newblock \showarticletitle{Designing Contestability: Interaction Design,
  Machine Learning, and Mental Health}. In
  \bibinfo{booktitle}{\emph{Proceedings of the 2017 Conference on Designing
  Interactive Systems}} \emph{(\bibinfo{series}{DIS '17})}.
  \bibinfo{publisher}{ACM}, \bibinfo{address}{New York, NY, USA},
  \bibinfo{pages}{95--99}.
\newblock
\showISBNx{978-1-4503-4922-2}
\urldef\tempurl%
\url{https://doi.org/10.1145/3064663.3064703}
\showDOI{\tempurl}


\bibitem[\protect\citeauthoryear{Miller, Howe, and Sonenberg}{Miller
  et~al\mbox{.}}{2017}]%
        {miller17}
\bibfield{author}{\bibinfo{person}{Tim Miller}, \bibinfo{person}{Piers Howe},
  {and} \bibinfo{person}{Liz Sonenberg}.} \bibinfo{year}{2017}\natexlab{}.
\newblock \showarticletitle{Explainable AI: Beware of Inmates Running the
  Asylum}. In \bibinfo{booktitle}{\emph{Proceedings of the 2017 IJCAI Workshop
  on Explainable Artificial Intelligence (XAI)}} \emph{(\bibinfo{series}{IJCAI
  '17})}. \bibinfo{pages}{36--42}.
\newblock
\showISBNx{978-1-4503-5970-2}
\urldef\tempurl%
\url{http://people.eng.unimelb.edu.au/tmiller/pubs/explanation-inmates.pdf}
\showURL{%
\tempurl}


\bibitem[\protect\citeauthoryear{Myers~West}{Myers~West}{2018}]%
        {MyersWest2018}
\bibfield{author}{\bibinfo{person}{Sarah Myers~West}.}
  \bibinfo{year}{2018}\natexlab{}.
\newblock \showarticletitle{Censored, suspended, shadowbanned: User
  interpretations of content moderation on social media platforms}.
\newblock \bibinfo{journal}{\emph{New Media {\&} Society}} \bibinfo{volume}{2},
  \bibinfo{number}{11} (\bibinfo{year}{2018}), \bibinfo{pages}{4366--4383}.
\newblock
\urldef\tempurl%
\url{https://doi.org/10.1177/1461444818773059}
\showDOI{\tempurl}


\bibitem[\protect\citeauthoryear{Office}{Office}{2019}]%
        {ParlEdu}
\bibfield{author}{\bibinfo{person}{Parliamentary~Education Office}.}
  \bibinfo{year}{2019}\natexlab{}.
\newblock \bibinfo{title}{Separation of Powers: Parliament, Executive and
  Judiciary}.
\newblock
\newblock
\urldef\tempurl%
\url{https://www.lawhandbook.org.au/2019_12_02_01_challenging_administrative_decisions/}
\showURL{%
Retrieved October 3, 2019 from \tempurl}


\bibitem[\protect\citeauthoryear{Ribeiro, Singh, and Guestrin}{Ribeiro
  et~al\mbox{.}}{2016}]%
        {Ribeiro16}
\bibfield{author}{\bibinfo{person}{Marco~Tulio Ribeiro},
  \bibinfo{person}{Sameer Singh}, {and} \bibinfo{person}{Carlos Guestrin}.}
  \bibinfo{year}{2016}\natexlab{}.
\newblock \showarticletitle{``Why Should I Trust You?": Explaining the
  Predictions of Any Classifier}. In \bibinfo{booktitle}{\emph{Proceedings of
  the 22nd ACM SIGKDD International Conference on Knowledge Discovery and Data
  Mining}} \emph{(\bibinfo{series}{KDD '16})}. \bibinfo{publisher}{ACM},
  \bibinfo{address}{New York, NY, USA}, \bibinfo{pages}{1135--1144}.
\newblock
\showISBNx{978-1-4503-4232-2}
\urldef\tempurl%
\url{https://doi.org/10.1145/2939672.2939778}
\showDOI{\tempurl}


\bibitem[\protect\citeauthoryear{Ribera and Lapedriza}{Ribera and
  Lapedriza}{2016}]%
        {Ribera19}
\bibfield{author}{\bibinfo{person}{Mireia Ribera} {and} \bibinfo{person}{Agata
  Lapedriza}.} \bibinfo{year}{2016}\natexlab{}.
\newblock \showarticletitle{Can we do better explanations? A proposal of
  User-Centred Explainable AI}. In \bibinfo{booktitle}{\emph{Joint Proceedings
  of the ACM IUI 2019 Workshop}}. \bibinfo{publisher}{ACM},
  \bibinfo{address}{New York, NY, USA}.
\newblock


\bibitem[\protect\citeauthoryear{Service}{Service}{2017}]%
        {Handbook}
\bibfield{author}{\bibinfo{person}{Fitzroy~Legal Service}.}
  \bibinfo{year}{2017}\natexlab{}.
\newblock \bibinfo{title}{Challenging administrative decisions}.
\newblock \bibinfo{howpublished}{The Law Handbook}.
\newblock
\urldef\tempurl%
\url{https://www.lawhandbook.org.au/2019_12_02_01_challenging_administrative_decisions/}
\showURL{%
Retrieved October 3, 2019 from \tempurl}


\bibitem[\protect\citeauthoryear{Vaccaro and Karahalios}{Vaccaro and
  Karahalios}{2017}]%
        {Vaccaro17}
\bibfield{author}{\bibinfo{person}{Kristen Vaccaro} {and}
  \bibinfo{person}{Karrie Karahalios}.} \bibinfo{year}{2017}\natexlab{}.
\newblock \showarticletitle{Algorithmic Appeals}. In
  \bibinfo{booktitle}{\emph{Workshop on Trustworthy Algorithmic
  Decision-Making}}.
\newblock
\urldef\tempurl%
\url{https://s3.amazonaws.com/kvaccaro.com/documents/algappeal.pdf}
\showURL{%
\tempurl}


\bibitem[\protect\citeauthoryear{Wachter, Mittelstadt, and Russell}{Wachter
  et~al\mbox{.}}{2018}]%
        {Watcher18}
\bibfield{author}{\bibinfo{person}{Sandra Wachter}, \bibinfo{person}{Brent
  Mittelstadt}, {and} \bibinfo{person}{Chris Russell}.}
  \bibinfo{year}{2018}\natexlab{}.
\newblock \showarticletitle{Counterfactual Explanations without Opening the
  Black Box: Automated Decisions and the GDPR}.
\newblock \bibinfo{journal}{\emph{Harvard Journal of Law and Technology}}
  \bibinfo{volume}{31}, \bibinfo{number}{2} (\bibinfo{year}{2018}),
  \bibinfo{pages}{841--887}.
\newblock


\bibitem[\protect\citeauthoryear{Wang, Yang, Abdul, and Lim}{Wang
  et~al\mbox{.}}{2019}]%
        {Wang19}
\bibfield{author}{\bibinfo{person}{Danding Wang}, \bibinfo{person}{Qian Yang},
  \bibinfo{person}{Ashraf Abdul}, {and} \bibinfo{person}{Brian~Y. Lim}.}
  \bibinfo{year}{2019}\natexlab{}.
\newblock \showarticletitle{Designing Theory-Driven User-Centric Explainable
  AI}. In \bibinfo{booktitle}{\emph{Proceedings of the 2019 CHI Conference on
  Human Factors in Computing Systems}} \emph{(\bibinfo{series}{CHI '19})}.
  \bibinfo{publisher}{ACM}, \bibinfo{address}{New York, NY, USA}, Article
  \bibinfo{articleno}{601}, \bibinfo{numpages}{15}~pages.
\newblock
\showISBNx{978-1-4503-5970-2}
\urldef\tempurl%
\url{https://doi.org/10.1145/3290605.3300831}
\showDOI{\tempurl}


\end{thebibliography}

\end{document}